# Binary topological logic gates in Kane–Mele nanostructures via local control of edge-state transport

K. Zberecki [a,*]

a Warsaw University of Technology, Faculty of Physics
* E-mail: krzysztof.zberecki@pw.edu.pl

## Abstract

Topological edge states offer a promising basis for post-CMOS device concepts. However, their use in elementary logic requires simple and physically transparent architectures. Here we study binary logic in Kane–Mele nanostructures with spatially localized control regions. Logical inputs are encoded by local electrostatic, exchange-like, and Rashba-type perturbations. The output is read from terminal transmission within the Landauer–Büttiker framework.
We demonstrate working NOT and AND gates in multiterminal honeycomb geometries. Real-space current maps show that their operation is governed by controlled rerouting of edge currents rather than by finely tuned interference. Both gates reproduce the complete truth table in all 50 Anderson-disorder realizations at each tested disorder strength up to $W/t = 0.10$. The NOT gate also remains correctly classified at all 441 points of a two-parameter control scan. Geometric tests confirm stable operation for several patch lengths and branch widths. These results establish Kane–Mele nanostructures as a transparent and numerically robust platform for primitive topological binary logic.

# 1. Introduction

The search for beyond-CMOS concepts for information processing at the micro- and nanoscale has stimulated sustained interest in topological quantum materials and nanostructures. In two-dimensional topological insulators, conducting boundary channels coexist with an insulating bulk, and in the quantum spin Hall regime these edge states are helical and protected by time-reversal symmetry against elastic backscattering from non-magnetic disorder [1–4]. The conceptual foundations of this field were laid by Kane and Mele in their seminal work on graphene [1,2], while the Bernevig–Hughes–Zhang framework and the subsequent HgTe experiments established a realistic route to the quantum spin Hall effect in solid-state nanostructures [5,6].
Over the past decade, the field has evolved from a largely conceptual one into a broader materials-and-device platform. Besides HgTe-based heterostructures [5,6], quantum spin Hall behavior has been identified or predicted in several atomically thin systems, including monolayer $WTe_2$, transition-metal dichalcogenide monolayers, and large-gap Kane–Mele-like platforms such as jacutingaite [7–9]. In parallel, the broader notion of topological electronics has gained momentum, with increasing attention devoted to whether protected boundary transport can be turned into robust device functionality at the micro- and nanoscale [10].
A central issue in this context is whether topological edge transport can be harnessed to perform elementary logic operations in a simple, physically transparent and reproducible way. A number of important studies have already demonstrated that transport through helical edge channels can be controlled by constrictions, gate-defined barriers, transistor-like switching concepts, Rashba-based spin manipulation, and field-effect tunneling between topological channels [11–15]. These works clearly establish that edge-state transport is controllable and device-relevant. However, they predominantly focus on switching, transistor action, or spin-current manipulation rather than on a unified implementation of standard binary logic gates in a microscopic lattice-based transport model. For topological nanoelectronics, there remains a clear need for architectures in which the logical inputs, the real-space current routing, and the output states can all be directly related to an intelligible transport mechanism, instead of relying primarily on narrow resonances or delicate interference conditions.

The Kane–Mele Hamiltonian provides a particularly attractive framework for such an investigation. As a paradigmatic lattice model of a two-dimensional topological insulator on the honeycomb lattice, it captures the essential physics of helical edge transport while remaining simple enough to support systematic device design and microscopic interpretation [1,2]. At the same time, its historical connection to graphene and its renewed relevance to modern large-gap and switchable quantum spin Hall materials make it especially suitable for nanoscale modeling of topological device concepts [2,8,9]. Within this framework, local electrostatic and exchange-like perturbations can be introduced in a controlled spatial manner, making it possible to selectively transmit, redirect, or suppress edge currents in multiterminal nanostructures.

In this work, we investigate binary logic in Kane–Mele nanostructures with spatially localized control regions. The basic idea is simple: logical inputs are encoded through local perturbations that change the accessibility of edge-state transport paths, while the output is read out from terminal-resolved transmission. Within this framework, Boolean functionality emerges directly from controlled current routing inside a single nanostructure. We show that NOT and AND operations can be realized in this way and identify operating windows in which the truth tables are reproduced together with clear transmission contrast and physically interpretable current maps. In the present context, the term topological logic refers to logic built on an edge-state transport backbone, even though the active control patches intentionally modify this backbone locally to redirect or suppress selected paths.

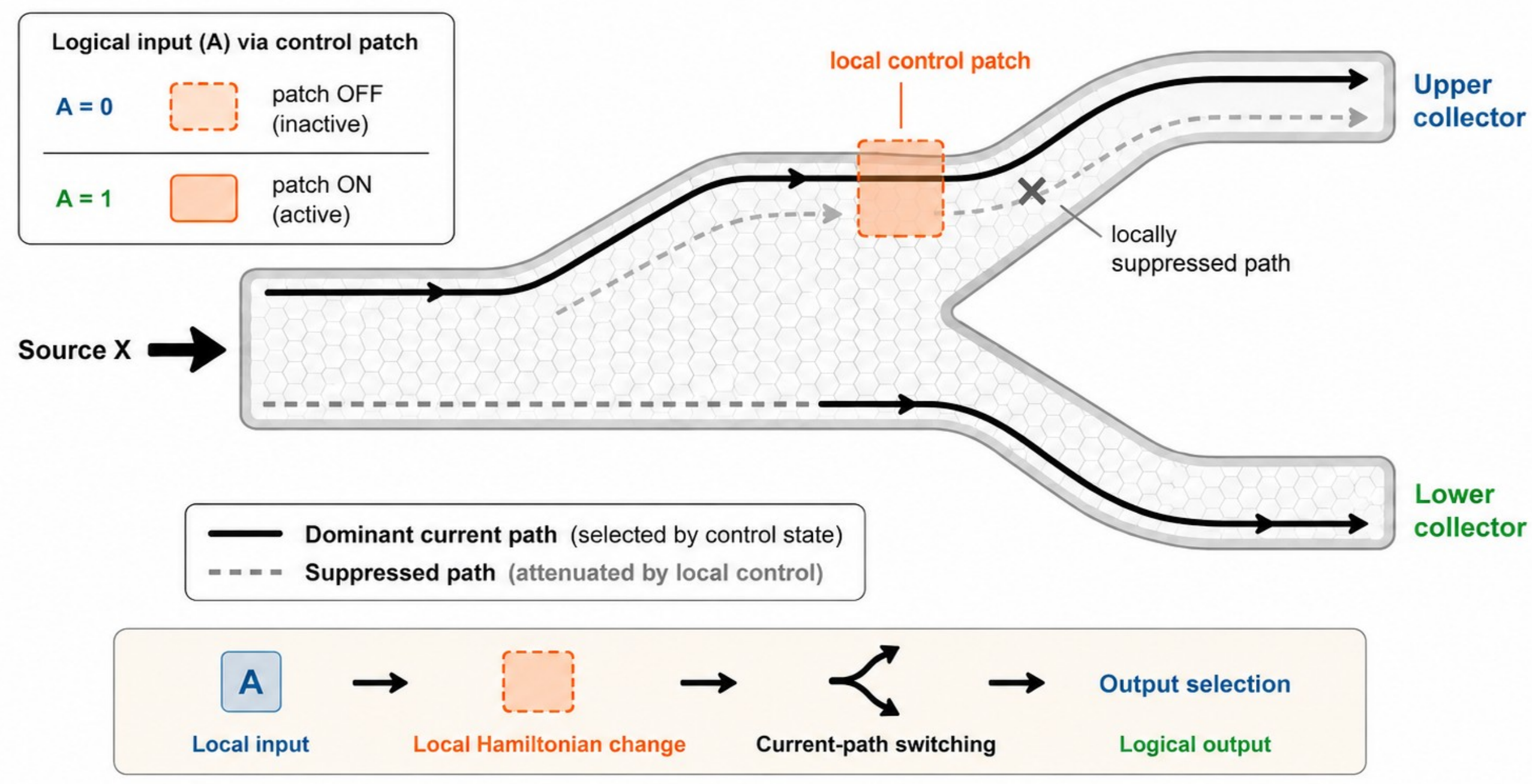

**Fig. 1** A localized control patch modifies the local Hamiltonian in a branched honeycomb device and redirects the dominant edge-current path between two complementary output collectors. The logical input is encoded through a prescribed local-control state, while the logical output is assigned from the collector receiving the dominant transmitted current. In this way, binary logic is realized by local control of edge-state transport rather than by finely tuned interference. This panel illustrates only the generic principle of current-path selection. The detailed input encoding and the assignment of logical values to the upper and lower collectors are gate-specific; in the NOT implementation of Fig. 3, the upper and lower branches represent Y=0 and Y=1, respectively, and the input selects between two complementary branch-control patches.

Although the original Kane–Mele proposal was formulated for graphene, whose intrinsic spin–orbit coupling is too weak for a straightforward electronic realization, closely related Kane–Mele-type physics has now been approached experimentally in several platforms. In particular, epitaxial bismuthene on SiC [16] provides a honeycomb-based large-gap quantum spin Hall candidate with spectroscopic signatures of conductive edge states, while monolayer $WTe_2$ [17] exhibits the quantum spin Hall effect up to 100 K [7]. More recently, low-buckled epitaxial germanene has been reported to display quantum spin Hall edge states together with an electrically driven topological phase transition [18] , which makes it especially relevant as a monoelemental material realization of Kane–Mele-type physics. These developments strengthen the case for using Kane–Mele nanostructures not only as a minimal theoretical framework, but also as a conceptually relevant platform for device-oriented topological logic. In the specific NOT implementation considered below, this generic routing principle is realized by activating one of two complementary branch-selective patches, rather than by simply switching a single spatial patch between its active and inactive states. The general logic concept studied here is illustrated schematically in Fig. 1. The essential idea is that a localized control patch modifies the local Hamiltonian inside a branched Kane–Mele nanostructure and thereby redirects the dominant edge-current path between complementary output branches. Binary input is thus encoded through a prescribed configuration of one or more fixed internal control regions, while binary output is read out from the collector receiving the dominant transmitted current. This generic mechanism underlies the more specific device geometries and gate implementations introduced below.

# 2. Theoretical model and methods

## 2.1. Kane–Mele Hamiltonian

We consider a multiterminal honeycomb nanostructure described by the tight-binding (TB) Kane–Mele Hamiltonian supplemented by local control terms. The total Hamiltonian is written as

$$H = H_{\mathrm{KM}} + H_{\mathrm{ctrl}} \tag{1}$$

The Kane–Mele part reads

$$H_{\mathrm{KM}} = -t \sum_{\langle i,j \rangle,\sigma} c_{i\sigma}^{\dagger} c_{j\sigma} + i\lambda_{\mathrm{SO}} \sum_{\langle\langle i,j \rangle\rangle,\alpha,\beta} \nu_{ij}\, c_{i\alpha}^{\dagger} (s_z)_{\alpha\beta} c_{j\beta} \tag{2}$$

The local control contribution is defined as

$$H_{\mathrm{ctrl}} = \sum_{i\in\Omega_V,\sigma} V_i\, c_{i\sigma}^{\dagger} c_{i\sigma} + \sum_{i\in\Omega_M,\alpha,\beta} c_{i\alpha}^{\dagger} (\mathbf{M}_i \cdot \mathbf{s})_{\alpha\beta} c_{i\beta} + i \sum_{\langle i,j \rangle \in\Omega_R,\alpha,\beta} \lambda_R^{ij}\, c_{i\alpha}^{\dagger} \left[ (\mathbf{s} \times \hat{\mathbf{d}}_{ij})_z \right]_{\alpha\beta} c_{j\beta} \tag{3}$$

Here, $c_{i\sigma}^{\dagger}$ and $c_{i\sigma}$ are the usual creation and annihilation operators for an electron with spin $\sigma$ on site $i$, $\mathbf{s} = (s_x, s_y, s_z)$ is the vector of Pauli matrices acting in spin space, and $\nu_{ij} = \pm 1$ is the Kane–Mele sign factor associated with the orientation of the next-nearest-neighbor hopping path. The sets $\Omega_V$, $\Omega_M$, and $\Omega_R$ denote the electrostatic, exchange-controlled, and Rashba-active regions, respectively. In a given device these regions may partly or fully overlap, so that a single control patch may carry electrostatic, exchange, and Rashba contributions simultaneously. The nearest- and next-nearest-neighbor bond sums are understood as oriented sums, with the hopping on the reverse bond generated by Hermitian conjugation, so that the full Hamiltonian remains Hermitian.

The parameter $t$ sets the basic energy scale of the model, while $\lambda_{\mathrm{SO}}$ is the intrinsic spin-orbit coupling acting on next-nearest-neighbor bonds and opening the topological gap that supports helical edge transport. The quantities $V_i$ represent local electrostatic on-site shifts applied inside prescribed control patches, whereas $\mathbf{M}_i$ denotes the corresponding local exchange field acting on the electron spin. When present, $\lambda_R^{ij}$ denotes the local Rashba spin-orbit coupling on nearest-neighbor bonds inside $\Omega_R$, and $\hat{\mathbf{d}}_{ij}$ is the unit vector pointing from site $j$ to site $i$. In the devices considered below, the Rashba term is used only as an additional local control contribution in selected gate regions and is taken to be piecewise constant inside the corresponding patch and zero elsewhere. In this sense, the local Rashba amplitude should be interpreted as an effective inversion-asymmetry contribution to the active patch Hamiltonian, rather than as a requirement that the Rashba interaction be switched independently from the other local control terms. For brevity, the symbol $\lambda_R$ will also be used below for the corresponding constant value of $\lambda_R^{ij}$ inside a given active control region. Unless stated otherwise, all energies and local-control amplitudes are expressed in units of the nearest-neighbor hopping $t$, so that $t = 1$ is used throughout, and the lattice constant is set to $a = 1$. With this convention, the parameter values quoted below, such as $\lambda_{\mathrm{SO}} = 0.06$ and the gate-dependent values of $V_i$, $\mathbf{M}_i$, and $\lambda_R$, should be understood as dimensionless quantities measured relative to the nearest-neighbor hopping.

## 2.2. Nanostructure geometry and terminals

All logic elements considered in this work are implemented in finite honeycomb nanostructures connected to semi-infinite leads of the same lattice type. The device geometries are designed such that transport in the relevant energy window is dominated by boundary channels of the Kane–Mele model, which provides a clear real-space basis for logic operation and current readout. In practice, this requires preserving well-defined outer boundaries and avoiding accidental geometric irregularities that could introduce spurious scattering centers or artificial conduction bottlenecks.

To ensure a transparent edge-dominated transport pattern, the scattering regions are constructed from zigzag-oriented honeycomb segments and junctions. This choice is motivated by the fact that, in the topological regime, zigzag-terminated ribbons provide a particularly clear visualization of boundary-guided current flow. In addition, the device shapes are generated in a row-resolved manner so as to avoid parasitic single-site protrusions or incomplete atomic rows at the interfaces between branches, constrictions and control regions. Such artifacts are not part of the intended device design and may otherwise obscure the interpretation of the transport mechanism.

The full nanostructure consists of three basic ingredients: input/output leads, a central scattering region, and spatially localized control patches. The leads act as ideal reservoirs and define the injection and readout channels of the logic device. The central region contains the junction or branching architecture that redistributes current between alternative boundary paths. The control patches are fixed spatial subregions in which local electrostatic and/or exchange-like perturbations can be applied in order to modify the accessibility of selected transport channels. In this way, logical inputs are encoded not by changing the leads themselves, but by switching the state of predefined internal regions of the nanostructure.

For the single-input gate geometries, the device is organized as a source lead feeding a branching region connected to two collector branches corresponding to the two possible logical outcomes. The role of the input-dependent control configuration is to determine which branch remains effectively connected to the incoming edge current. For the two-input gate geometries, the same general idea is extended by introducing two independently controlled stages, labelled $\Omega_A$ and $\Omega_B$, each of which may contain more than one fixed patch location and acts sequentially or cooperatively on the available transport paths. The logical output is then inferred from the transmission to a designated output terminal, while auxiliary branches may serve as complementary collectors or current drains depending on the specific gate design.

This geometry-first construction is central to the present work. Rather than relying on fine tuning of resonant states, the proposed logic elements are based on spatially controlled redirection, blocking, or preservation of boundary currents inside a multiterminal honeycomb nanostructure. As a result, the logic functionality can be interpreted directly in terms of the real-space connectivity of edge-state transport paths, which is advantageous both conceptually and numerically. Specific gate-dependent dimensions, branch widths, and control-region placements are introduced in the corresponding subsections below.

The generic device blueprint and notation used throughout this work are summarized in Fig. 2, while the common Hamiltonian and operating parameters for the NOT and AND gate implementations are collected in Table 1. The complete numerical definitions of the reference NOT and AND geometries, including the splitter positions, branch widths, interbranch gaps, and control-patch boundaries, are listed in Supplementary Table S1.

The specific operating points and device dimensions used below were not chosen arbitrarily, but were identified through systematic transport scans combined with iterative geometry refinement. Since the present devices are proof-of-principle lattice geometries rather than optimized lithographic layouts, we do not assign a unique universal minimum size to the schematic blueprint in Fig. 2. Instead, the relevant geometric dimensions are specified for each gate and their influence on the logical margin is assessed through the size-scaling tests summarized in Sec. 3.4 and Supplementary Fig. S1.

In practice, candidate structures and parameter sets were evaluated using three joint criteria: exact reproduction of the target truth table, positive transmission margins for all tested logical input states, and physically interpretable real-space current maps consistent with edge-guided transport. The working points reported in the following sections should therefore be understood as representative solutions selected from a broader design-search procedure rather than as isolated ad hoc parameter choices.

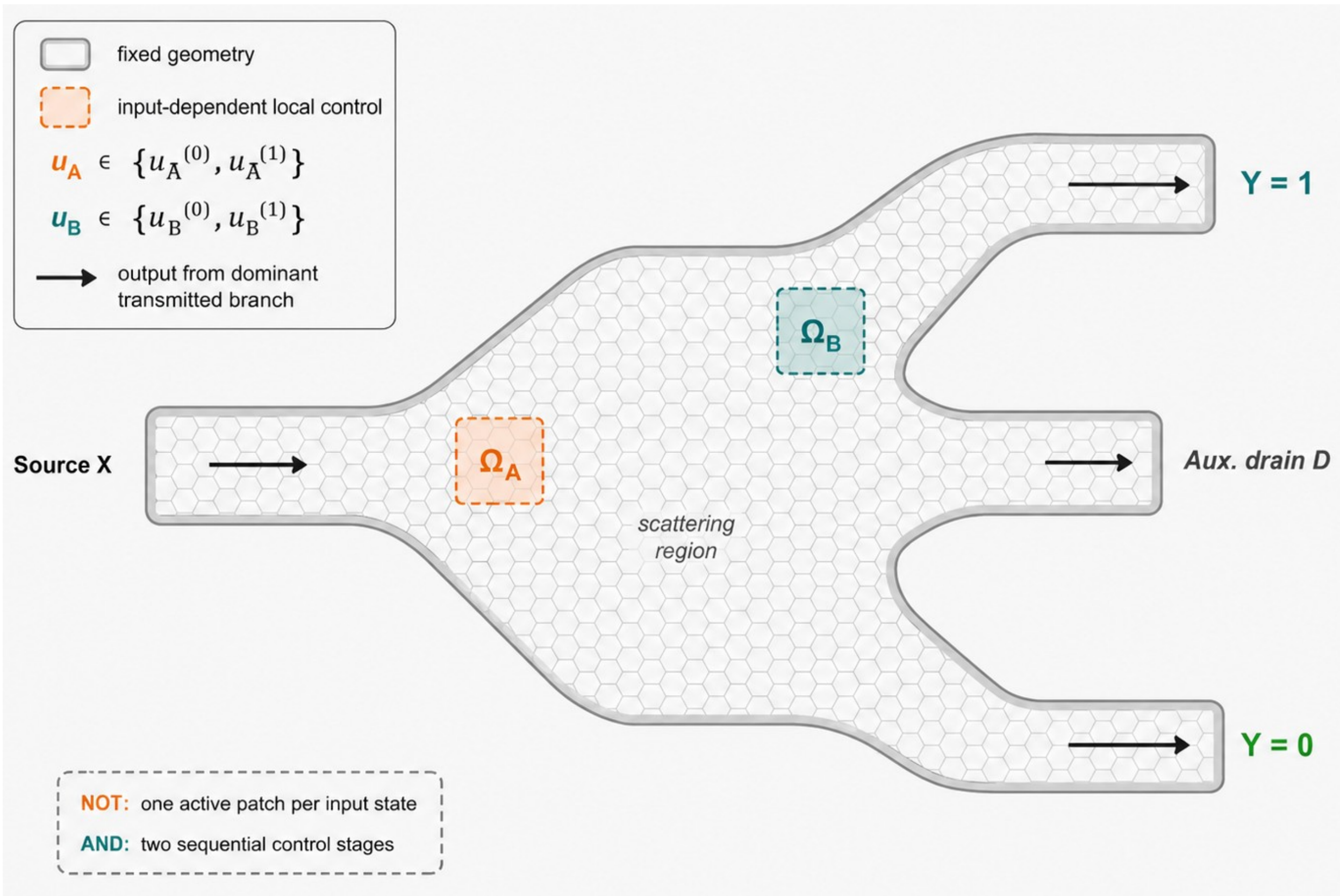

**Fig. 2** Generic device blueprint and notation for the Kane–Mele nanostructure family used for binary topological logic. The common multiterminal platform consists of a honeycomb-lattice scattering region connected to a source lead $X$, complementary logical collectors $Y=1$ and $Y=0$, and an optional auxiliary drain $D$. The vertical placement of the $Y=1$ and $Y=0$ labels in this generic blueprint is schematic and does not prescribe the upper/lower collector assignment in the gate-specific geometries. Logical inputs are encoded through spatially fixed local control patches, denoted $\Omega_A$ and $\Omega_B$, whose Hamiltonian parameters are switched between discrete states without changing the device geometry. The logical output is assigned from the branch receiving the dominant transmitted current. This common notation underlies both the complementary-patch NOT device, in which one logical input selects between upper and lower branch-control regions, and the two-stage AND architecture. The figure is intended as a schematic notation-level blueprint; the gate-specific dimensions, control-patch lengths and branch widths are defined in the corresponding device geometries and tested explicitly in the size-scaling analysis reported in Supplementary Fig. S1.

| [illegible] | [illegible] | [illegible] | [illegible] |
|---|---|---|---|
| Nearest-neighbor hopping | $t$ | 1.0 | 1.0 |
| Intrinsic spin–orbit coupling | $\lambda_{SO}$ | 0.06 | 0.06 |
| Operating energy | $E$ | 0.0 | -0.010 |
| Input structure | — | single input A | two inputs A and B |
| Logical output readout | — | dominant transmission into $Y=1$ or $Y=0$ branch | comparison between transmission into the $Y=1$ branch and the summed complementary $Y=0$ sector |
| Control architecture | — | two complementary upper/lower branch-selective patches; exactly one patch active for each input state | two sequential stages, each containing a complementary pair of branch-selective patches |
| Electrostatic control amplitude | $V$ | $V_{\text{block}}=0.20$ | $V_A$ = 0.20; $V_B$ = 0.02 |
| Exchange-field amplitude (along x) | $M_x$ | $M_{x,\text{block}}=0.45$ | $M_{A,x}$ = 0.45; $M_{B,x}$ = 0.23 |
| Local Rashba coupling | $\lambda_R$ | $\lambda_{R,\text{block}}=0.22$ | $\lambda_{R,A}$ = 0.22; $\lambda_{R,B}$ = 0.04 |
| Representative geometry | — | branching splitter with complementary upper and lower branch-control patches | two-stage branched device with stage-A and stage-B regions |
| Physical operating principle | — | input-controlled selection of complementary branch-blocking patches | sequential path selection; only $(A,B)=(1,1)$ yields $Y=1$ |

**Table 1** Common Hamiltonian and operating parameters used in the simulations for the binary NOT and AND gates. The table summarizes the model parameters and gate-specific local control amplitudes at the selected operating points.

## 2.3. Quantum-transport framework

Transport through the proposed Kane–Mele nanostructures is analyzed within the Landauer–Büttiker scattering framework, which is well suited for multiterminal mesoscopic devices in the linear-response regime [19,20]. In this approach, the semi-infinite leads act as ideal reservoirs supporting propagating modes, while the finite device region defines the scattering problem. For a given Fermi energy $E$, the logical response is characterized primarily through terminal-resolved transmission probabilities between the input and output leads. This representation is especially convenient here because the logical state can be read out directly from the transmitted signal without introducing additional phenomenological assumptions.
The main transport quantity is the transmission from lead $i$ to lead $j$, denoted $T_{j\leftarrow i}(E)$. In the scattering-matrix formalism, it is given by

$$\mathrm{T}_{j\leftarrow i}(E) = \mathrm{Tr}\left[t_{ji}^{\dagger}(E)\, t_{ji}(E)\right] \tag{4}$$

or, equivalently, as a sum over incoming and outgoing propagating modes,

$$T_{j\leftarrow i}(E) = \sum_{n\in i}\sum_{m\in j} |S_{mn}(E)|^2 \,. \tag{5}$$

Because the transmission coefficient is defined as a sum over propagating modes, its value is not restricted to the interval $[0,1]$ in the multichannel regime. In particular, terminal-resolved transmissions and their sums may exceed unity when more than one incoming mode is available in the source lead. In this work, the reported dimensionless transmission coefficients should therefore be interpreted as mode-summed observables in the standard Landauer–Büttiker sense, rather than as single-channel probabilities. Accordingly, throughout the manuscript quantities such as $T_{Y=1\leftarrow X}$, $T_{Y=0\leftarrow X}$, and $R_{\mathrm{in}}$ denote mode-summed terminal observables, not normalized single-particle probabilities.
Throughout this work, we use the dimensionless transmission itself as the primary logic observable. When needed, it can be related to the corresponding linear-response conductance through the Landauer formula,

$$G_{j\leftarrow i}(E) = \frac{e^2}{h}\, T_{j\leftarrow i}(E). \tag{6}$$

In addition to terminal transmissions, we analyze the spatial distribution of current inside the scattering region in order to identify the transport mechanism responsible for a given logical response. This is done by evaluating bond-current expectation values for scattering states injected from a chosen source lead. Real-space current maps provide a direct way to determine whether the device operates through boundary-guided transport, through redistribution of current between competing branches, or through local suppression of selected paths by the control regions. In this sense, they are not merely illustrative, but serve as an important diagnostic of the physical origin of the logic action.
For a scattering state incident from lead $i$, the current on the bond connecting sites $m$ and $n$ is computed from the corresponding tight-binding current operator,

$$\hat{J}_{m\to n} = \frac{i}{\hbar}\left(\mathbf{c}_m^{\dagger} H_{mn}\mathbf{c}_n - \mathbf{c}_n^{\dagger} H_{nm}\mathbf{c}_m\right), \tag{7}$$

where $H_{mn}$ is the hopping matrix element between the two sites and $c_n$ denotes the local spinor annihilation operator. For a given scattering state $|\psi^{(\alpha)}\rangle$, the corresponding bond current is evaluated as

$$J_{m\to n}^{(\alpha)} = \frac{2}{\hbar}\,\mathrm{Im}\left[\psi_m^{(\alpha)\dagger} H_{mn}\psi_n^{(\alpha)}\right]. \tag{8}$$

In the numerical implementation, both the scattering matrix and the local current operator are evaluated using the Kwant package [21] , which provides a natural framework for mode-resolved quantum transport in multiterminal tight-binding systems. This allows us to combine terminal readout with real-space diagnostics in a unified and internally consistent way. In the figures shown below, the displayed current maps are obtained by summing the bond-current expectation values over all propagating modes incident from the source lead at the chosen energy, with equal unit-flux weighting.

To further assess whether the relevant transport remains edge-dominated, we use an edge-weight diagnostic constructed from the mode-summed density of the scattering states injected from the source lead. For a given logical input configuration, let $\mathcal{M}_X(E)$ denote the set of propagating modes incident from the source lead $X$ at energy $E$, and let $\psi_i^{(\alpha)\dagger}(E)\psi_i^{(\alpha)}(E)$ be the spin-summed local density of the unit-flux-normalized scattering state associated with incoming mode $\alpha$. The edge weight is then defined as

$$f_{\mathrm{edge}}(E) = \frac{\displaystyle\sum_{\alpha\in\mathcal{M}*X(E)} \sum *i \in \Omega_{\mathrm{edge}} \rho_i^{(\alpha)}(E)}{\displaystyle\sum_{\alpha\in\mathcal{M}*X(E)} \sum *i \in \Omega_{\mathrm{dev}} \rho_i^{(\alpha)}(E)}. \tag{9}$$

Here, $\Omega_{\mathrm{dev}}$ contains all sites of the finite scattering region, while $\Omega_{\mathrm{edge}}$ is constructed from the nearest-neighbor connectivity of the device. Boundary sites are identified as sites with reduced nearest-neighbor coordination relative to the maximum coordination in the device. For the results reported below, the edge region contains the boundary sites together with the first inward nearest-neighbor shell, corresponding to two graph shells in total. All propagating source modes are included with equal unit-flux weighting.

Although this quantity is not itself used for logical readout, it is useful for confirming that the operating regime remains associated with edge-state transport rather than with bulk-like conduction through the interior of the device. This distinction is important for interpreting the gate mechanism and for judging whether the resulting functionality is genuinely topological in character. Unless stated otherwise, all transport characteristics are evaluated at fixed energy inside the spectral window where the finite-width nanoribbons retain the topological edge-transport character of the Kane–Mele model.

## 2.4. Logic encoding and output readout

The logical inputs of the device are encoded through discrete states of predefined local control regions embedded in the scattering geometry. For a binary gate, each input variable takes values in the set $A, B \in \{0, 1\}$.

In the single-input case, only one such variable is active, whereas in the two-input case two independently controlled patch configurations, associated with $A$ and $B$, are used. Each logical value corresponds to a prescribed configuration of local control parameters, such as electrostatic amplitudes, exchange-field strengths, or combinations thereof, applied inside one or more fixed spatial patches of the nanostructure.

To keep the notation general, we denote by $u_\xi$ the control parameter associated with input ξ, where $\xi = A$ or $B$. The binary encoding is then written as

$$u_\xi(\xi) = u_\xi^{(0)}(1-\xi) + u_\xi^{(1)}\xi, \qquad \xi \in \{A, B\}, \tag{10}$$

where $u_\xi^{(0)}$ and $u_\xi^{(1)}$ denote the parameter values corresponding to the logical states $0$ and $1$, respectively. In this way, the geometric location of a control region remains fixed, while its local Hamiltonian contribution changes according to the logical input. For architectures containing more than one fixed control patch, $u_A^{(0)}$ and $u_A^{(1)}$ denote complete input-dependent patch configurations rather than two values of a single scalar parameter. This convention is used for the NOT gate below, where both complementary patch locations remain fixed and exactly one of them is active for each input state.
The full transport problem for a given input configuration is defined by the input-dependent Hamiltonian

$$H(A,B) = H_{\mathrm{KM}} + H_{\mathrm{ctrl}}[u_A(A), u_B(B)] \,. \tag{11}$$

For single-input gates, the same notation applies after suppressing the inactive variable. This formulation is convenient because it makes the logical state of the device equivalent to a discrete choice of local Hamiltonian parameters, while the underlying nanostructure geometry and lead configuration remain unchanged.
The logical output is extracted from terminal-resolved transmission. In the geometry considered here, a source lead $X$ injects current into the device, while two designated collector leads correspond to the logical outputs $Y = 1$ and $Y = 0$. We define

$$T_1(E;A,B) = T_{Y_1 \leftarrow X}(E;A,B), \qquad T_0(E;A,B) = T_{Y_0 \leftarrow X}(E;A,B). \tag{12}$$

If an additional auxiliary drain is present, its transmission is monitored separately as

$$T_{\mathrm{aux}}(E;A,B) = T_{D \leftarrow X}(E;A,B), \tag{13}$$

but it is not itself interpreted as a logical state. Instead, it serves as a measure of parasitic leakage or of current removed from the logical output channels.
In geometries with more complex branching, one logical output may be represented by a complementary sector composed of more than one physical collector, in which case the corresponding logical transmission is defined as the sum of the relevant terminal-resolved contributions. This convention is used for the AND gate below, where the logical ($Y = 0$) output is represented by the sum of the transmissions into the middle and lower complementary collectors.
Unless stated otherwise, the binary output is assigned from the dominant logical-output transmission,

$$Y(A,B) = \underset{y \in \{0,1\}}{\arg\max}\, T_y(E;A,B). \tag{14}$$

This criterion is natural for multiterminal devices with complementary logical outputs, because it identifies the logical state with the physical branch or summed output sector carrying the larger transmitted signal. In geometries where only one designated logical-output branch is used, the same readout can be expressed in threshold form,

$$\text{Y(A,B)} = \begin{cases} 1, & T_1(E;A,B) \geq T_{\mathrm{th}}, \\ 0, & T_1(E;A,B) < T_{\mathrm{th}}, \end{cases} \tag{15}$$

where $T_{th}$ is an appropriately chosen decision threshold. In the present work, the dominant-output criterion is the primary one, while gate-specific thresholds may be used as auxiliary diagnostics when needed.
To quantify the quality of the logical response, we compare the assigned output with the expected truth-table value $Y_{exp}(A,B)$. A useful measure of the logical contrast for a given input configuration is the transmission margin

$$m(A,B) = T_{Y_{\mathrm{exp}}}(E;A,B) - T_{1-Y_{\mathrm{exp}}}(E;A,B), \tag{16}$$

which is positive when the transmission assigned to the correct logical output exceeds that assigned to the incorrect logical output. If an auxiliary drain is present, its role is analyzed separately through $T_{aux}(E; A, B)$; a large positive margin together with low leakage provides the clearest signature of well-defined gate operation.
At the level of the full truth table, the logical performance can be summarized by the exact classification rate

$$\eta = \frac{1}{N_{\text{cases}}} \sum_{c=1}^{N_{\text{cases}}} \delta_{Y_c,\, Y_c^{\text{exp}}}, \tag{17}$$

where $N_{cases}$ is the number of tested input combinations, $Y_c$ is the assigned output for case c, and $Y_c^{exp}$ is the corresponding expected output. In addition, we monitor the set of margins $m(A, B)$ in order to distinguish merely correct truth-table reproduction from more robust logical discrimination. This distinction becomes particularly important when the device is later tested against moderate variations of energy and control parameters.
For completeness, we also monitor the input reflection,

$$R_{\text{in}}(A, B) = R_{X \leftarrow X}(E; A, B), \tag{18}$$

which measures the mode-summed reflection back into the source lead. In the multichannel regime, $R_{\text{in}}$ should therefore be interpreted on the same footing as the mode-summed terminal transmissions, rather than as a fraction restricted to the interval $[0, 1]$.
Although input reflection is not itself used as the logical decision variable, it provides an important auxiliary measure of transport efficiency. Since it is evaluated as a mode-summed reflection back into the source lead, values larger than unity simply indicate that more than one incoming source mode contributes to the reflected signal. This quantity becomes particularly relevant when assessing the suitability of a given gate design for future cascaded implementations.

# 3. Results and discussion

## 3.1. Binary NOT gate

We first consider the realization of a binary NOT gate in the nanostructure shown in Fig. 3. The device is a single-input branching Kane–Mele nanostructure with one source lead and two collector branches associated with the complementary logical outputs $Y = 0$ and $Y = 1$. Two spatially fixed branch-selective control patches are placed in the upper and lower output arms. The single logical input $A$ determines which of these two complementary patches is activated, while the other patch remains inactive. The logical output is assigned from the branch receiving the larger terminal transmission.

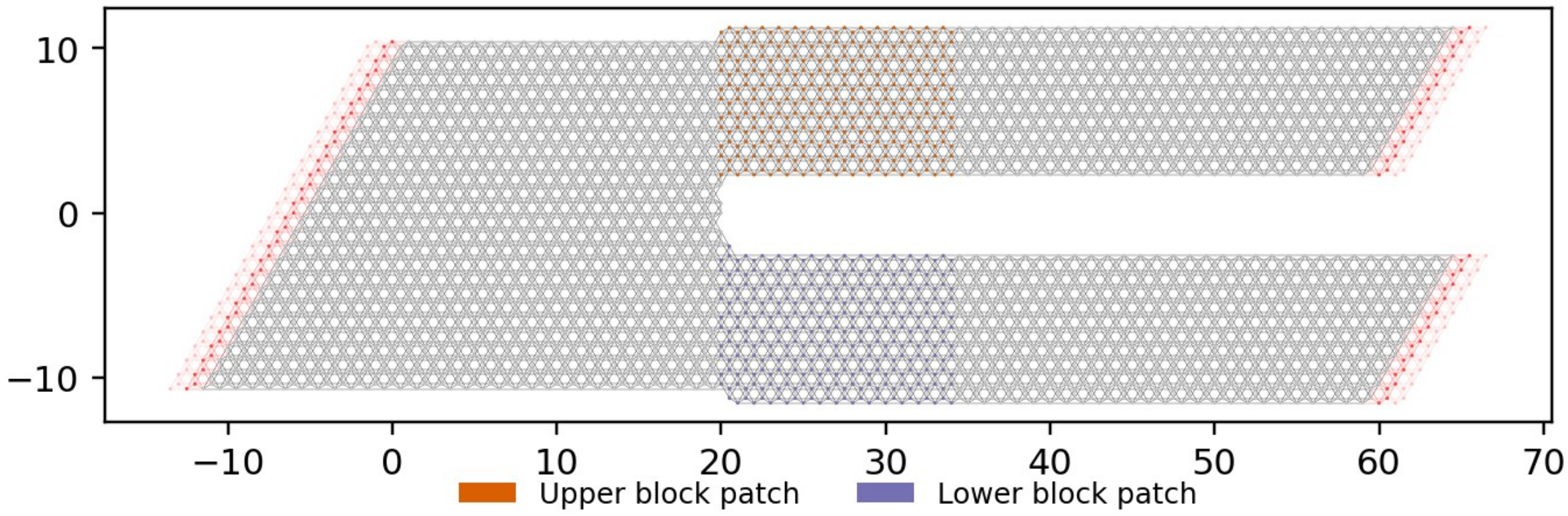

**Fig. 3** Kane–Mele nanostructure used for the binary NOT gate. The honeycomb-lattice scattering region is connected to one source lead and two complementary output branches. The highlighted upper and lower regions are branch-selective control patches associated with the $Y = 0$ and $Y = 1$ paths, respectively. For $A = 0$, the upper patch is activated and the current is redirected toward the lower $Y = 1$ branch. For $A = 1$, the lower patch is activated and the dominant current exits through the upper $Y = 0$ branch. Exactly one of the two complementary patches is active for each logical input state.

The input encoding is defined as follows. For $A = 0$, the upper control patch is activated and suppresses transmission into the upper $Y = 0$ branch, thereby redirecting the current toward the lower $Y = 1$ collector. For $A = 1$, the lower control patch is activated and suppresses the lower $Y = 1$ branch, so that the dominant current exits through the upper $Y = 0$ collector. Thus, the two input states use the same local-control amplitudes but apply them to complementary spatial regions. At the operating point used below, the active patch is characterized by an electrostatic contribution $V_{\text{block}} = 0.20$, a local exchange field $M_{x,\text{block}} = 0.45$, and a Rashba contribution $\lambda_{R,\text{block}} = 0.22$. The remaining parameters are fixed to $t = 1.0$, $\lambda_{\text{SO}} = 0.06$, and $E = 0$.

The operating principle of the device is illustrated by the local current maps in Fig. 4. For $A = 0$, the upper branch-selective patch is active. Its combined electrostatic, exchange, and Rashba contributions suppress the direct path into the upper $Y = 0$ collector, and the injected edge current is redirected toward the lower branch, yielding $Y = 1$. For $A = 1$, the input configuration activates the lower branch-selective patch instead. The lower $Y = 1$ path is then suppressed, while the current preferentially reaches the upper collector, yielding $Y = 0$. The gate action therefore corresponds to input-controlled switching between two complementary spatial perturbations, which invert the preferred current path without changing the underlying device geometry. In the current maps, the quantity shown as ``current magnitude" is the absolute value of the locally interpolated bond-current field, constructed from the Kwant current operator acting on scattering states incident from the source lead.

To support the real-space interpretation quantitatively, we also evaluated the edge-weight diagnostic introduced in Sec. 2.3. For both input states of the NOT gate, the corresponding scattering states remain strongly edge-localized, with $\mathrm{f}_{\text{edge}} = 0.950$ for A=0 and $\mathrm{f}_{\text{edge}} = 0.957$ for A=1. This confirms that the gate action is built on an edge-dominated transport backbone rather than on bulk-like conduction through the interior of the device.

This behavior is summarized in Table 2. For $A = 0$, the transmission into the branch representing $Y = 1$ is close to unity, $T_{Y=1\leftarrow X} = 1.087690$, while the complementary output remains subdominant. For $A = 1$, the transmission into the same branch drops to $T_{Y=1\leftarrow X}$=0.164189, whereas the transmission into the $Y = 0$ branch increases to $T_{Y=0\leftarrow X} = 1.108428$. According to the readout rule introduced in Sec. 2.4, this corresponds to the truth table

$$A = 0 \ \rightarrow \ Y = 1, \qquad A = 1 \ \rightarrow \ Y = 0, \tag{19}$$

which is precisely the desired NOT operation.

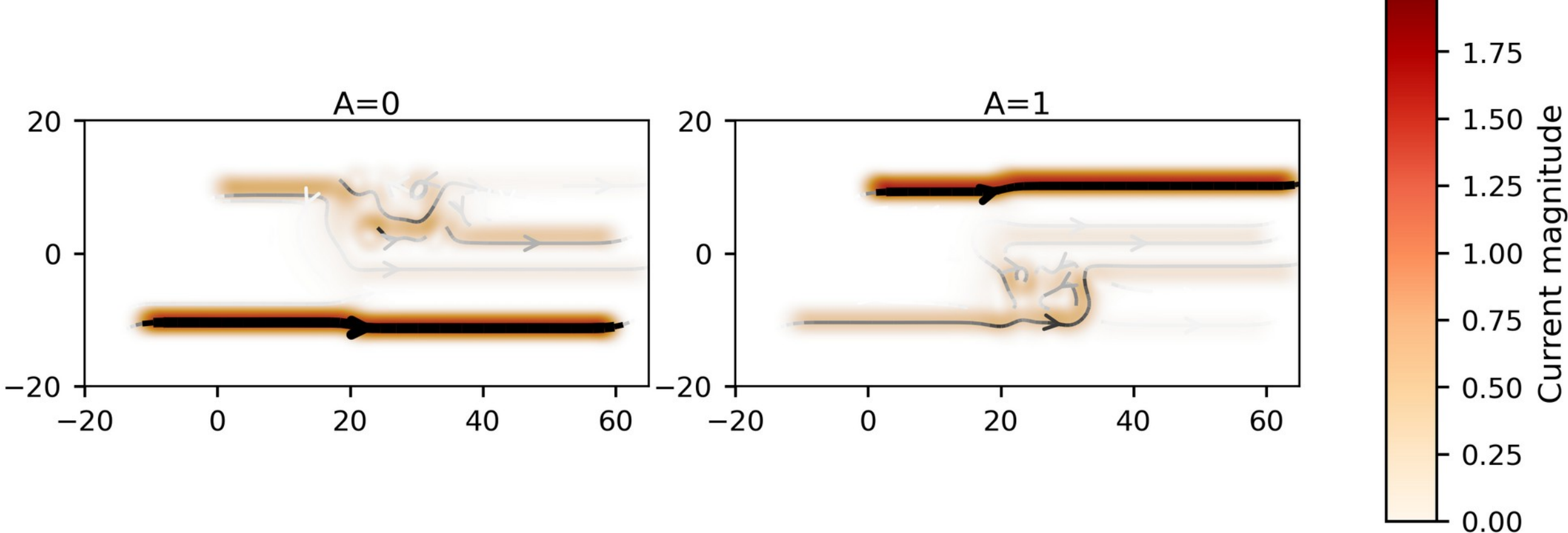

**Fig. 4** Real-space current maps for the Kane–Mele NOT gate for the two binary input states. For $A = 0$, the upper branch-selective patch is active, suppressing the $Y = 0$ path and redirecting the dominant current toward the lower $Y = 1$ branch. For $A = 1$, the lower patch is active, suppressing the $Y = 1$ path and yielding the upper-branch output $Y = 0$. This complementary branch switching provides a direct real-space signature of the NOT operation. The black arrows mark the dominant edge-current path.

An important aspect of this result is that the logic functionality does not arise from a narrowly tuned interference minimum, but from a spatially local and physically interpretable reorganization of edge-state transport. The logical input selects which of the two complementary branch-selective patches modifies the local connectivity of the edge-current path. Activation of the upper patch suppresses the $Y = 0$ branch and produces $Y = 1$, whereas activation of the lower patch suppresses the $Y = 1$ branch and produces $Y = 0$. This distinction is clearly visible in the real-space current maps and is consistent with the strong transmission contrast between the two input states. The NOT gate therefore provides a transparent demonstration of how binary inversion can be encoded through spatial selection of local control regions and read out from edge-current routing.

| | | |
|---|---|---|
| A | 0 | 1 |
| Expected Y | 1 | 0 |
| Assigned Y | 1 | 0 |
| T(Y=0 ← X) | 0.296958 | 1.108428 |
| T(Y=1 ← X) | 1.087690 | 0.164189 |
| $R_{in}$ | 0.615352 | 0.727383 |
| Margin | 0.790732 | 0.944239 |
| Status | PASS | PASS |

**Table 2** Truth table and transmission-based readout for the binary NOT gate at the selected operating point. For each input state A, the table reports the expected logical output, the output assigned from the transmission-based readout, the transmissions into the $Y = 0$ and $Y = 1$ branches, the input reflection $R_{in}$, the logical margin, and the final classification status.

## 3.2. Binary AND gate

The two-input AND gate is realized within a Kane–Mele nanodevice in which two independently controlled routing stages act sequentially on the accessible transport paths. In contrast to the complementary-patch NOT gate discussed above, the AND architecture contains two pairs of fixed branch-selective control patches. Stage A contains upper and lower patches controlled by the input $A$, whereas stage $B$ contains two downstream patches controlled by the input $B$. For each input value, exactly one patch within the corresponding pair is active. The two-stage geometry is shown in Fig. 5. The two stages are therefore not arranged as two geometrically equivalent input arms. Instead, the AND geometry is deliberately asymmetric: stage A acts as an upstream sector selector, whereas stage B is placed farther downstream and acts as a final discriminator between the $Y = 1$ collector and the complementary $Y = 0$ sector.

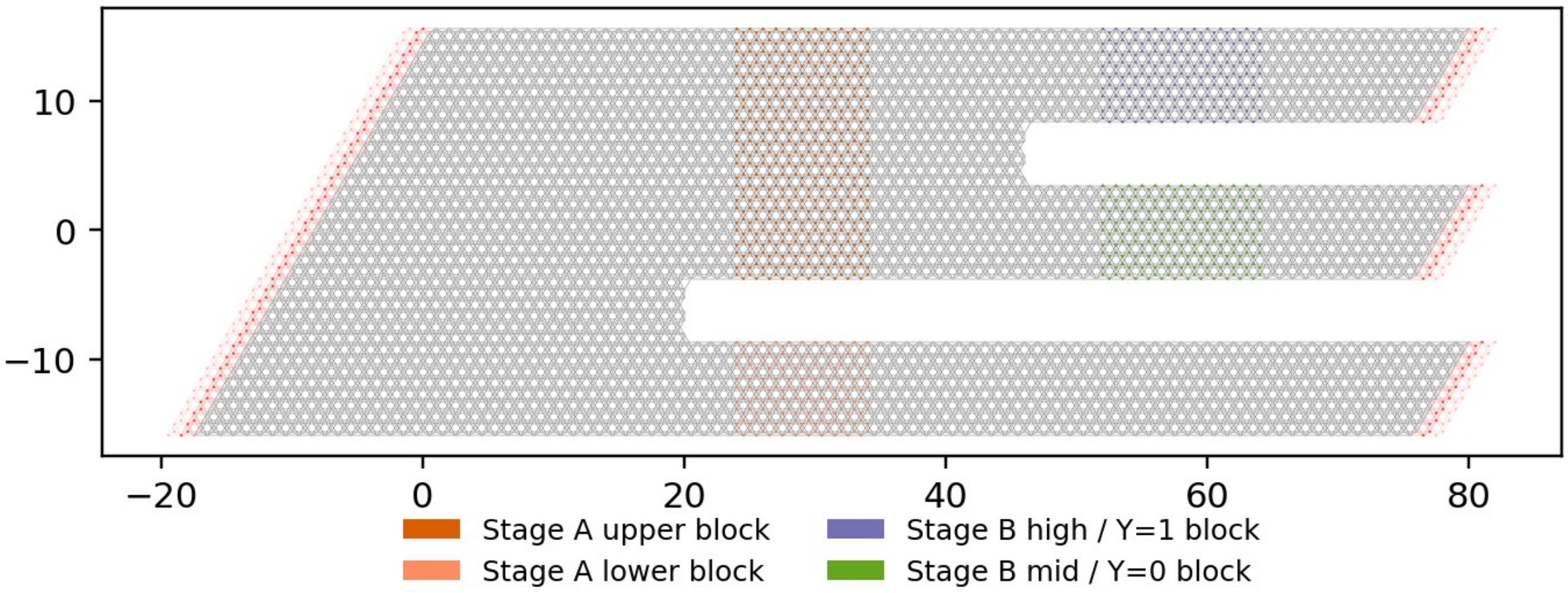

**Fig. 5** Geometry of the Kane–Mele nanostructure used for the binary AND gate. The highlighted internal regions mark the functional control patches: stage-A upper and lower blocking zones (orange shades) and stage-B regions associated with the $Y = 1$ collector and the complementary $Y = 0$ sector (blue and green, respectively). The device is intentionally asymmetric: stage A acts as an upstream selector that controls access to the upper transport sector, whereas stage B is placed farther downstream and performs the final discrimination between the $Y = 1$ branch and the complementary $Y = 0$ output sector. The unequal branch lengths are therefore part of the sequential AND design and are used to spatially separate the two control functions. Specifically, $A = 0$ and $A = 1$ activate the stage-A upper and lower patches, respectively, whereas $B = 0$ and $B = 1$ activate the stage-B high-$Y = 1$ and middle-$Y = 0$ blocking patches, respectively.

The input-to-patch mapping is defined as follows:

| Input state | Active blocking patch |
|---|---|
| A = 0 | $\Omega_{A,\text{upper}}$ |
| A = 1 | $\Omega_{A,\text{lower}}$ |
| B = 0 | $\Omega_{B,\text{high}}$ |
| B = 1 | $\Omega_{B,\text{mid}}$ |

For each logical input value, the complementary patch belonging to the same stage remains inactive. The stage-A upper patch suppresses access to the upper transport sector, whereas the stage-A lower patch suppresses the lower complementary branch. Within stage B, the high patch suppresses the $Y = 1$ rail, while the middle patch suppresses the upper $Y = 0$ rail. Thus, the logical inputs select which transport path is locally blocked rather than directly switching a selected output path on.

The resulting sequential routing mechanism is straightforward. For $A = 0$, the stage-A upper patch is active and prevents efficient access to the upper part of the device. The current is therefore directed toward the lower complementary $Y = 0$ collector, irrespective of the value of B . For $A = 1$, the stage-A lower patch is active, suppressing the lower branch and allowing the current to enter the upper transport sector. The final output is then selected by stage B. For B=0 , the stage-B high patch suppresses the Y=1 rail and redirects the current into the complementary Y=0 sector, predominantly through the middle collector. For $B=1$, the stage-B middle patch suppresses the upper complementary $Y=0$ path, leaving the high Y=1 collector as the dominant output. Consequently, a continuous path to $Y=1$ exists only for $(A,B)=(1,1)$, which produces the AND truth table through sequential branch blocking.

In the working implementation considered here, the logical inputs $A$ and $B$ select one active patch within each of the two complementary patch pairs associated with stages $A$ and $B$, respectively. The operating point is chosen at $E=-0.010$, $t=1.0$, and $\lambda_{\mathrm{SO}}=0.06$. For the first stage, the active control patch is characterized by $V_A=0.20$, $M_{x,A}=0.45$, and $\lambda_{R,A}=0.22$, whereas the second stage is governed by the weaker downstream perturbation $V_B=0.02$, $M_{x,B}=0.23$, and $\lambda_{R,B}=0.04$. Within each stage, the same control amplitudes are assigned to the two complementary patch locations; the corresponding logical input determines which of these locations carries the active perturbation.

This hierarchy of amplitudes is physically meaningful: the first stage performs the stronger selection of the transport sector, while the weaker downstream stage acts on the already preconditioned current flow and discriminates between the final logical branches without introducing excessive backscattering or spurious current loops.

The device geometry was refined to support this two-stage mechanism in a stable manner. In particular, the splitter leading to the upper logical channel was made more adiabatic, and the spatial separation between the upper and lower sectors was increased so as to reduce unintended current leakage and recirculation. The resulting branch lengths are not meant to define a left-right or upper-lower symmetric structure. Rather, they reflect the sequential logic architecture: the stage-A control patch occupies the central part of the structure and acts on the current before it reaches the upper branching region, while the stage-B patch is placed farther downstream near the upper logical collector. This spatial offset separates the upstream selection step from the downstream discrimination step. If the two control stages are placed too close to one another, or if the branches are insufficiently isolated, the intended sequential action is partially lost and the current distribution becomes less clearly associated with the logical truth table.

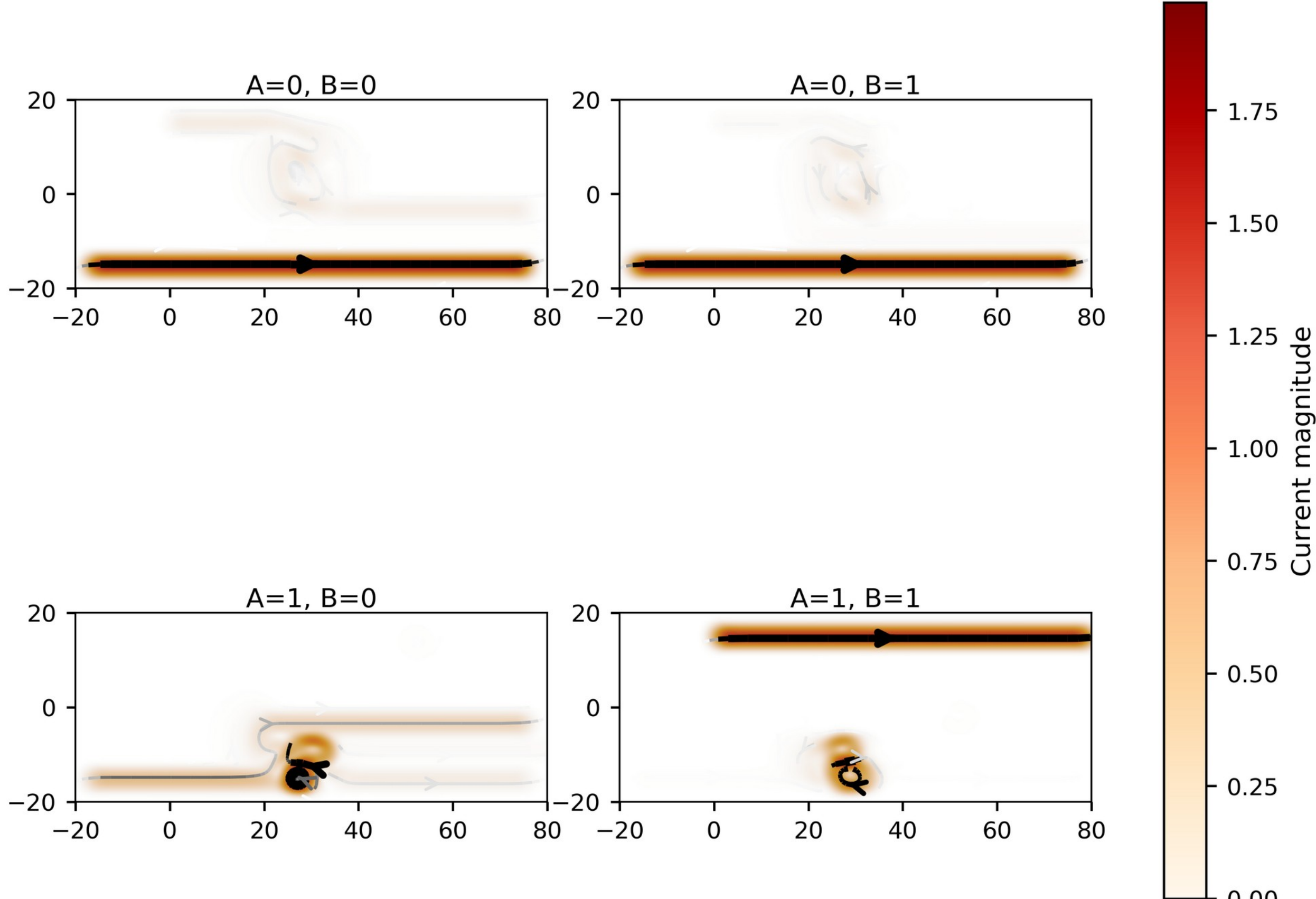

**Fig. 6** Local current maps for the binary AND gate at the selected operating point. The four panels correspond to the input combinations $(0,0)$, $(0,1)$, $(1,0)$, and $(1,1)$. Only the $(1,1)$ configuration supports a continuous transmitting path toward the $Y=1$ collector, whereas the remaining three cases are redirected into the complementary $Y=0$ sector. The maps show local current-density patterns, whereas the logical assignment is made from terminal-integrated transmissions; partial local current penetration into a branch is therefore not sufficient to define the logical output.

The current maps in Fig. 6 illustrate the physical operation of the gate for the four binary input combinations. For $(A,B)=(0,0)$, $(0,1)$, and $(1,0)$, the current does not establish an efficient terminal path toward the Y=1 collector and is instead redirected into the complementary $Y=0$ sector. These three configurations therefore yield the logical output $Y=0$. By contrast, for $(A,B)=(1,1)$, the two routing stages act cooperatively and a continuous edge-guided path is formed through the upper sector of the device toward the $Y=1$ collector.
It is important to distinguish local current penetration into a branch from the terminal-integrated logical readout. In particular, the $(A,B)=(1,0)$ configuration shows some current density in the upper part of the device, but this current does not form an efficient transmitting path to the $Y=1$ terminal. The summed transmission into the complementary $Y=0$ sector therefore remains dominant, and the state is correctly classified as $Y=0$. Thus, only the $(1,1)$ configuration exhibits a continuous transmitting path to the upper logical collector, whereas the remaining three cases show incomplete access to this path and enhanced current diversion into the complementary output sector.
The edge-weight diagnostic further confirms that the relevant operating states of the AND device remain predominantly edge-dominated, with $f_{\mathrm{edge}}=0.974$, $0.974$, $0.832$, and $0.932$ for the four input combinations $(0,0)$, $(0,1)$, $(1,0)$, and $(1,1)$, respectively. The smallest value occurs for the $(A,B)=(1,0)$ configuration, which is also the state with the largest input reflection and the smallest logical margin; nevertheless, even in this case the transport remains clearly edge-dominated rather than bulk-like.

The resulting logical response is summarized by the standard AND truth table,

$$(A,B) \in \{(0,0),(0,1),(1,0)\} \;\rightarrow\; Y=0, \qquad (1,1) \rightarrow Y=1. \tag{20}$$

In the present geometry, the logical output $Y=0$ is represented by two physical branches. Accordingly, we define the total transmission into the $Y=0$ sector as

$$T_{Y=0,\mathrm{total}}(A,B) = T_{Y=0,\mathrm{mid}\leftarrow X}(A,B) + T_{Y=0,\mathrm{low}\leftarrow X}(A,B). \tag{21}$$

As shown in Table 3, the selected operating point reproduces this truth table correctly for all four input combinations. Importantly, the distinction between the logical states is not merely formal: it is accompanied by clear differences in the terminal-resolved transmissions and by positive logical margins for the correctly classified cases. This confirms that the AND gate is not obtained through marginal threshold crossings, but through a robust redistribution of current between physically distinct output channels. The most delicate case is $(A,B)=(1,0)$, for which the current map shows partial access to the upper sector. Nevertheless, the terminal transmission into the $Y=1$ collector remains negligible, $T_{Y=1\leftarrow X}=0.001490$, whereas the summed complementary transmission is $T_{Y=0,\mathrm{total}\leftarrow X}=0.661561$. The resulting positive margin, $m=0.660071$, confirms that this state is correctly classified as $Y=0$. At the same time, the transmission data show that correct logical classification does not always coincide with optimal impedance matching or minimal back-reflection. In particular, some input configurations still exhibit substantial reflection into the source lead, even though the dominant-output criterion and the logical margin remain correct. The present AND element should therefore be viewed primarily as a proof-of-principle primitive gate at the transport level rather than as a fully optimized building block for direct multistage cascading.

| | | | | |
|---|---|---|---|---|
| A | 0 | 0 | 1 | 1 |
| B | 0 | 1 | 0 | 1 |
| Expected Y | 0 | 0 | 0 | 1 |
| Assigned Y | 0 | 0 | 0 | 1 |
| T(Y=1 ← X) | 0.000003 | 0.002146 | 0.001490 | 1.000906 |
| T(Y=0_mid ← X) | 0.082669 | 0.000603 | 0.427158 | 0.000942 |
| T(Y=0_low ← X) | 1.005434 | 1.021501 | 0.234403 | 0.005387 |
| T(Y=0_total ← X) | 1.088103 | 1.022104 | 0.661561 | 0.006328 |
| $R_{in}$ | 0.911894 | 0.975750 | 1.336949 | 0.992766 |
| Margin | 1.088101 | 1.019958 | 0.660071 | 0.994578 |
| Status | PASS | PASS | PASS | PASS |

**Table 3** Truth table and transmission-based readout for the binary AND gate at the selected operating point. For each input pair $(A,B)$, the table reports the expected logical output, the assigned output obtained from the transport readout, the transmission into the $Y=1$ branch, the partial transmissions into the two branches forming the complementary $Y=0$ sector, their total contribution $T(Y=0_{\mathrm{total}} \leftarrow X)$, the input reflection $R_{\mathrm{in}}$, the logical margin, and the final classification status.

Together with the current maps, these results show that the AND operation is realized through sequential control of edge-current routing in a multiterminal Kane–Mele nanostructure.

## 3.3. Additional binary gates

The direct transport-level realizations of the NOT and AND gates obtained above are sufficient to generate the remaining standard binary logic operations by composition, since the set $\{\mathrm{NOT}, \mathrm{AND}\}$ is functionally complete [22] . In particular,

$$\begin{aligned} \mathrm{NAND}(A,B) &= \neg\big(A \wedge B\big), \\ \mathrm{OR}(A,B) &= \neg\big(\neg A \wedge \neg B\big), \\ \mathrm{NOR}(A,B) &= \neg\big(A \vee B\big), \\ \mathrm{XOR}(A,B) &= (A \vee B) \wedge \neg(A \wedge B), \\ \mathrm{XNOR}(A,B) &= \neg\big(A \oplus B\big). \end{aligned} \tag{22}$$

Thus, the primitive NOT and AND elements demonstrated here already provide the minimal building blocks required for constructing the full binary gate family at the Boolean-design level. However, this statement should not be interpreted as a full transport-level demonstration of directly cascaded multistage logic. In the present Landauer–Büttiker setting, the output of a gate is a terminal-resolved transmission or current contrast between physical collector leads, not a restored voltage logic level with intrinsic gain. Therefore, a practical cascaded architecture would require additional circuit-level elements or device-engineering steps, including current-to-voltage conversion, signal restoration, impedance matching between successive stages, fan-out control, and suppression of accumulated reflection and leakage. The constructions in Eq. (22) should consequently be understood as Boolean compositions of validated primitive transport gates, while the realization of fully cascaded topological logic circuits remains an important direction for future work.

## 3.4. Stability and robustness

To assess the robustness of the proposed topological logic functionality, we examined the response of the two primitive gates established in this work, namely the NOT and AND elements, using three complementary tests: static Anderson-type disorder, moderate variations of the operating-point parameters, and systematic variations of selected geometric dimensions around the reference device layouts. These tests distinguish numerically stable logic devices from solutions that reproduce the correct truth table only at a narrowly tuned parameter point or for a single particular geometry.

We first tested the stability of the two gates against static Anderson-type on-site disorder. For each site (i) in the finite scattering region, we added a random scalar contribution $\delta V_i s_0$ to the local Hamiltonian, where $\delta V_i$ was drawn independently from a uniform distribution in the interval $[-W/2, W/2]$. The disorder was applied only inside the finite scattering region, while all semi-infinite leads were kept clean. For each disorder strength $W$, we generated 50 independent disorder realizations. For a given realization, the same disorder configuration was used for all logical input states of the corresponding gate, so that the full truth table was evaluated for one and the same disordered device.

The results of this disorder ensemble are summarized in Fig. 7. The tested disorder strengths were $W/t = 0, 0.02, 0.05, 0.075$, and $0.10$. For both the NOT and AND gates, the full truth-table pass rate remains equal to unity over the entire tested range, Fig. 7(a). In other words, all 50 disorder realizations reproduce the correct truth table for both primitive gates at every disorder strength considered.

The logical margins, shown in Fig. 7(b), provide a more stringent measure of robustness. The NOT gate retains large positive margins throughout the scan. Its mean minimum margin remains close to $0.8$ $W = 0.10t$ $0.8$, and even at the strongest tested disorder, $W = 0.10t$, the worst realization still has a minimum margin of approximately $0.54$. This confirms that the single-stage NOT architecture is well separated from the decision boundary and remains robust against moderate static on-site disorder.

The AND gate is also correctly classified in all 50 disorder realizations for all tested W, but its margins are more strongly affected by disorder. The mean minimum margin decreases from about $0.66$ in the clean case to about $0.23$ at $W = 0.10t$. The worst minimum margin at the strongest disorder is still positive, approximately $0.02$, but it is much smaller than for the NOT gate. This behavior is physically consistent with the more complex two-stage architecture of the AND element, in which the final logical output depends on preserving a preconditioned transport path through two spatially separated control regions. Thus, while the AND gate remains fully correct within the tested disorder window, it is naturally more sensitive to disorder than the simpler NOT gate.

A complementary operating-window analysis was performed by varying selected control amplitudes around the operating points; the corresponding minimum-margin maps are shown in Fig. 8. For the NOT gate, we performed a two-dimensional sensitivity scan in the plane of the electrostatic patch amplitude $V_{\text{block}}$ and the exchange amplitude $M_{x,\text{block}}$, while keeping the remaining parameters fixed at the reference values $E = 0$, $\lambda_{\text{SO}} = 0.06t$, and $\lambda_R = 0.22t$. For each point of the $21 \times 21$ grid, the full NOT truth table was recalculated and the minimum logical margin over the two input states was extracted. The NOT gate reproduces the correct truth table at all $441$ scanned points in the range $0.10 \leq V_{\text{block}}/t \leq 0.30$ and $0.30 \leq M_{x,\text{block}}/t \leq 0.60$. The minimum margin remains positive throughout the scanned window, with the smallest value approximately $0.02$. The reference point used in the truth-table and current-map calculations, $V_{\text{block}} = 0.20t$ and $(M_{x,\text{block}} = 0.45t$, is therefore not an isolated fine-tuned solution but lies inside a finite region of correct NOT operation.

For the AND gate, Fig. 8b shows the operating-window map in the $(E, V_A)$ plane. The tested values were $E/t = -0.015, -0.010, -0.005$ and $V_A/t = 0.15, 0.20, 0.25$. For each $(E, V_A)$ point, the full AND truth table was evaluated for $V_B/t = 0.00, 0.02, 0.04$, and the smallest logical margin over both the four input configurations and the three tested $V_B$ values was retained. The remaining local-control parameters were fixed at $M_{x,A}/t = 0.45$, $\lambda_{R,A}/t = 0.22$, $M_{x,B}/t = 0.23$, and $\lambda_{R,B}/t = 0.04$. The AND gate remains correctly classified throughout the scanned window, although its minimum margins are smaller than those of the NOT gate. This behavior is consistent with the sequential two-stage architecture of the AND element and with its stronger disorder sensitivity observed in Fig. 7b.

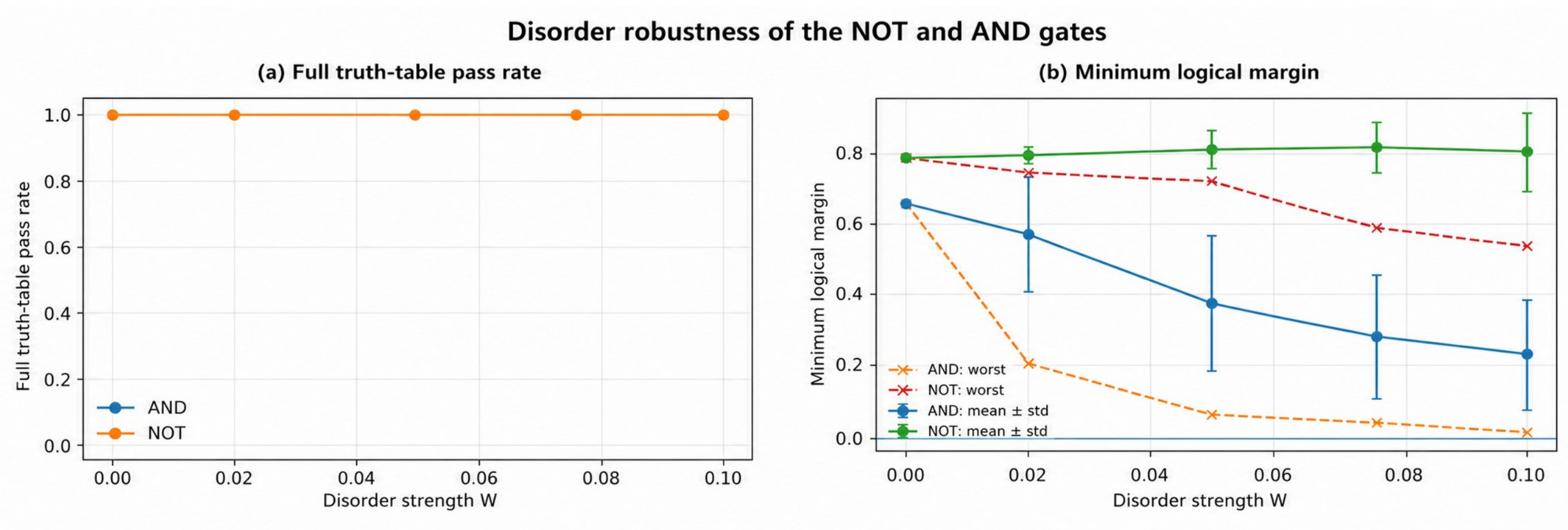

**Fig. 7** Disorder robustness of the primitive NOT and AND gates under Anderson-type on-site disorder. For each disorder strength $W$, 50 independent disorder realizations were generated. The same disorder configuration was used for all logical input states within a given realization, and the full truth table was tested. **(a)** Full truth-table pass rate as a function of ($W$). Both gates retain a pass rate equal to unity up to ($W/t = 0.10$). The NOT and AND pass-rate curves coincide exactly in panel (a), so the two traces are superimposed. **(b)** Minimum logical margin over the truth table,

shown as the mean $\pm$ standard deviation over disorder realizations together with the worst realization. The NOT gate remains well separated from the decision boundary, while the AND gate shows a systematic margin reduction but remains positive throughout the tested range.

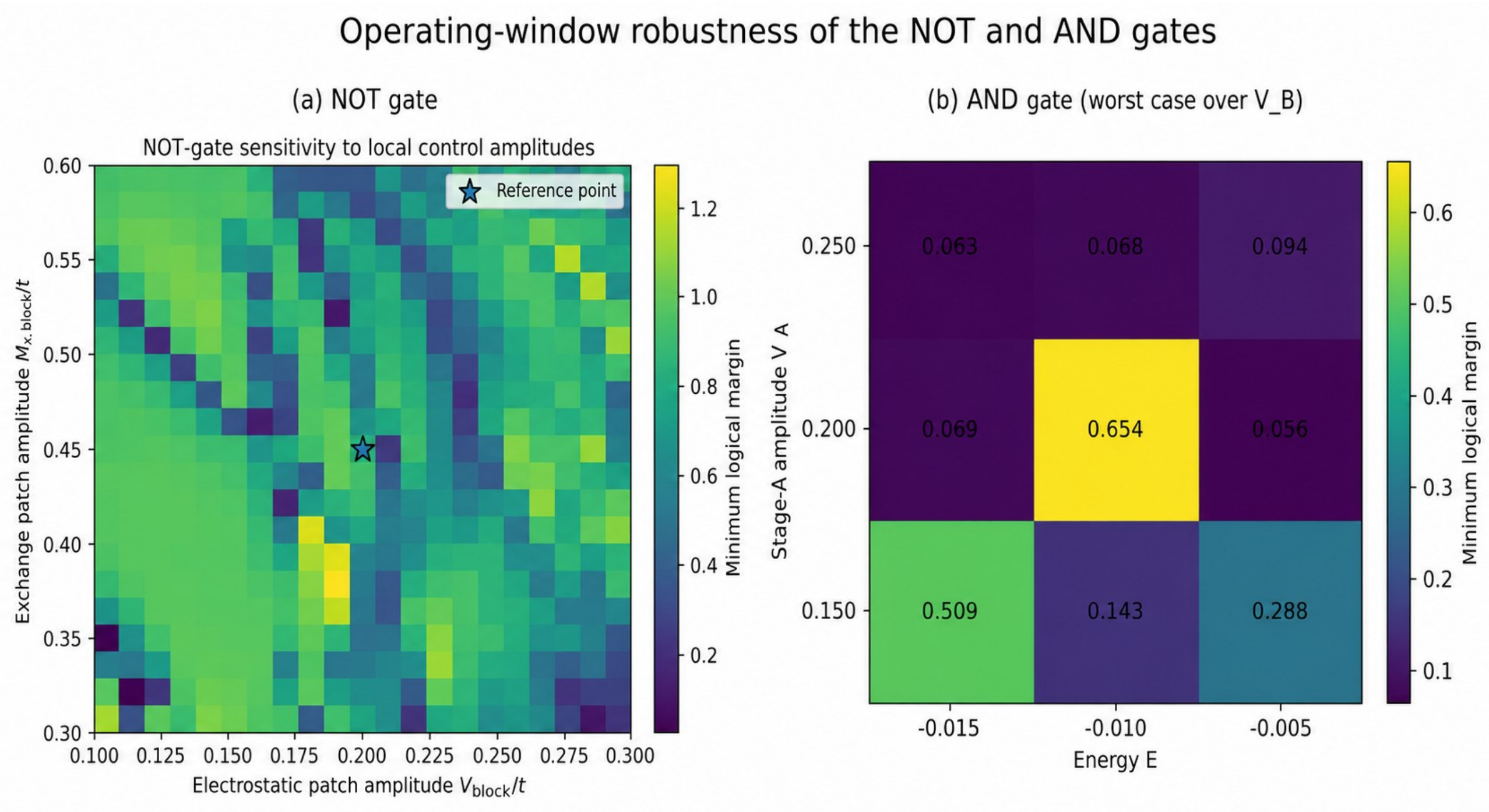

**Fig. 8** Operating-window robustness of the primitive NOT and AND gates. (a) Minimum logical margin for the NOT gate as a function of the electrostatic patch amplitude $V_{\mathrm{block}}$ and the exchange amplitude $M_{x,\mathrm{block}}$. The star marks the reference operating point used in the truth-table and current-map calculations. All 441 scanned points reproduce the correct NOT truth table, and the minimum margin remains positive throughout the plotted window. (b) Worst-case minimum logical margin for the AND gate as a function of the operating energy $E$ and the stage-A electrostatic amplitude $V_A$. At each $(E, V_A)$ point, the minimum was taken over the complete AND truth table and over the tested stage-B amplitudes $V_B/t = 0.00, 0.02, 0.04$. Positive margins indicate correct logical classification. The two panels use independent color scales because the characteristic margin ranges of the NOT and AND gates differ substantially.

Finally, we examined how the logical margins depend on characteristic geometric dimensions of the devices. Rather than attempting to define an atomically sharp universal minimum size, which would depend on the discrete row structure of a particular honeycomb realization, we varied the main control-patch lengths and branch widths around the reference geometries. For the NOT gate, the common longitudinal extent of the upper and lower complementary control patches was varied while their centers and local-control amplitudes were kept fixed. For each logical input state, exactly one of the two patches was active. The NOT gate remains correctly classified over the full tested range of the common patch length, while its branch width is more restrictive. In the present row-resolved geometry, stable operation with a positive margin is obtained for several sufficiently wide and row-compatible branches, including the reference width $W_{\mathrm{branch}} = 10$. The AND gate also remains correctly classified over all tested patch lengths and branch widths. Its logical margin is most sensitive to the stage-A patch length, whereas the stage-B patch length and the final subbranch width have only a weak influence over the tested range. These results indicate that the reference geometries are not isolated size-specific solutions, although the AND stage-A geometry and the NOT branch width require more careful geometric design than the common longitudinal extent of the complementary NOT patch pair. Additional geometric size-scaling data are provided in Supplementary Fig. S1.

Taken together, these tests show that the proposed topological binary gates are not isolated numerical solutions. The disorder ensemble demonstrates that both primitive gates reproduce the full truth table in all 50 independent realizations up to $W/t = 0.10$. The NOT sensitivity map further shows that the inverter remains correctly classified over a broad $(V_{\mathrm{block}}, M_{x,\mathrm{block}})$ window, while the AND operating-window scan confirms correct operation in the tested parameter range, albeit with smaller margins. The geometric size-scaling analysis additionally shows that both reference devices remain functional over multiple control-patch lengths and branch widths, although the NOT branch width and the AND stage-A geometry require more careful design. The combined robustness analysis therefore supports the central claim of this work, namely that local control of edge-state transport in Kane–Mele nanostructures can yield primitive binary logic elements whose operation is both physically transparent and numerically stable.
The present robustness analysis is intentionally restricted to static scalar on-site disorder, selected operating-point variations, and regular changes of several characteristic patch lengths and branch widths around the reference geometries. It does not include more general fabrication imperfections, such as edge roughness, vacancies or reconstructed boundaries, irregular control-patch shapes, random patch displacements, correlated spatial disorder, finite-temperature broadening, or dephasing. These effects remain outside the scope of the present proof-of-principle study and should be addressed in future work.

## 3.5. General design principles for topological binary logic

The results obtained above suggest that binary logic in Kane–Mele nanostructures should be designed primarily as a problem of controlled real-space current routing rather than as a search for isolated transmission resonances. In the present devices, the logical response is determined by whether an edge-current path remains open, is redirected into a complementary branch, or is suppressed by a localized perturbation. This picture leads to a few practical design rules:
1. The ungated nanostructure should already support a clear edge-dominated transport sector. In practice, this requires well-defined outer boundaries, smooth branch junctions, and the elimination of geometric artifacts such as incomplete rows or parasitic single-site protrusions.
2. Local perturbations should act on specific transport paths at specific stages of the device. In the NOT gate, a single logical input selects between two complementary branch-selective patches, with exactly one patch active in each input state, whereas the AND gate requires two spatially separated control stages with distinct functions.
3. Electrostatic, exchange, and Rashba-like terms are most effective when they act cooperatively but not uniformly. In particular, the AND device benefits from a stronger upstream perturbation that prepares the transport sector and a weaker downstream perturbation that performs the final branch discrimination.
4. The logical outputs should correspond to physically distinct collectors whose transmissions can be compared directly. Excessive coupling between neighboring branches increases leakage, reduces the logical margin, and makes the readout more sensitive to disorder and parameter drift.
5. A useful topological logic element should not only reproduce the correct truth table, but also maintain positive logical margins under moderate disorder and operating-point variations. In this respect, the NOT gate retains larger logical margins throughout its tested control-parameter scan, whereas the more complex AND gate remains functional but exhibits smaller margins and a stronger sensitivity to disorder.
From an experimental perspective, the nanostructures considered here are not purely abstract. Recent roadmap assessments for two-dimensional topological insulators [23] explicitly identify electrically controlled edge transport, heterostructure engineering with magnetic or superconducting layers, and device-level functionality as realistic directions for near-term development. In particular, gate-defined weak links have already been fabricated in monolayer $WTe_2$ [24], showing that locally gated junction-like regions can be sculpted directly in a quantum spin Hall material. In parallel, $WTe_2/Cr_2Ge_2Te_6$ heterostructures demonstrate that exchange-like perturbations can be induced by magnetic proximity and detected in edge-dominated transport [25] , while STM/STS experiments on bismuthene show that bringing topological edges into close proximity can partially lift their protection through inter-edge hybridization [26] . Taken together, these advances suggest that branched nanostructures of the type studied here could, in principle, be approached experimentally by combining patterned or gate-defined quantum-spin-Hall channels with local electrostatic gates and, where needed, magnetic-proximity regions that emulate the model exchange patches. A conceptually even closer long-term route may be offered by proximity-engineered graphene platforms, for which quantum spin Hall states at zero magnetic field have recently been reported [27] .
The Rashba term used in the model should be viewed in the same effective-device sense. Microscopically, a local Rashba contribution requires structural inversion asymmetry and may be produced or enhanced by a perpendicular electric field, an asymmetric dielectric or substrate environment, interface engineering, or proximity to a material with strong spin-orbit coupling. In a practical implementation, the logical state of a control patch need not correspond to the independent real-time switching of the Rashba coupling alone. Rather, it may represent a combined local Hamiltonian configuration in which electrostatic gating provides the fastest tunable component, while exchange and inversion-asymmetry contributions are supplied by the device stack or by more slowly configurable local environments. The present simulations therefore demonstrate the transport functionality of such effective local control regions, while the optimization of a specific high-speed experimental switching protocol remains a separate device-engineering problem.
Taken together, these observations suggest a general design strategy for topological binary logic in Kane–Mele nanostructures: begin from a clean edge-transport geometry, assign distinct logical roles to spatially localized control patches, use perturbations hierarchically rather than uniformly, isolate the competing output branches, and evaluate success through robust transmission margins rather than truth-table matching alone. Within this framework, the primitive NOT and AND elements established here provide more than isolated examples. They define a set of operational rules for constructing binary logic directly from controlled edge-state transport, and they offer a practical starting point for future multistage topological logic architectures based on Boolean composition of physically validated gate primitives.

# 4. Conclusions

We have proposed and analyzed a nanoscale binary-logic architecture based on Kane–Mele nanostructures with spatially localized control regions. Within a unified quantum-transport framework, we identified working implementations of the primitive NOT and AND gates and showed that their truth tables can be reproduced at selected operating points together with clear transmission contrast and physically interpretable current maps.
In both devices, the logical response is governed by controlled reconfiguration of edge-state transport paths rather than by narrowly tuned interference. The NOT gate realizes this mechanism through input-controlled selection between two complementary branch-blocking patches, whereas the AND gate extends it to a sequential two-stage architecture. The disorder analysis shows that both gates reproduce their complete truth tables in all 50 independent realizations at every tested disorder strength up to $W/t = 0.10$. The NOT element remains correctly classified at all 441 points of the scanned $(V_{\text{block}}, M_{x,\text{block}})$ parameter window, while the AND element retains positive margins at all points of the tested $(E, V_A, V_B)$ parameter grid, although its margins are more strongly reduced by disorder.
The geometric size-scaling analysis further demonstrates that the reported operating points are not tied to single, isolated device dimensions. Correct logical classification is retained over multiple control-patch lengths and branch widths, although the NOT branch width and the AND stage-A geometry emerge as the more restrictive design parameters. Taken together, these results support

Kane–Mele nanostructures as a transparent platform for primitive topological binary logic. Since the set $(\mathrm{NOT}, \mathrm{AND})$ is functionally complete, the demonstrated elements provide the minimal building blocks required for constructing the remaining binary gate family at the Boolean-design level. A practical transport-level cascading architecture would nevertheless require current-to-voltage conversion, signal restoration, impedance matching, fan-out control, and further suppression of reflection and leakage. Future work should also address edge roughness, irregular patch placement, finite-temperature broadening, dephasing, and experimentally motivated implementations in quantum-spin-Hall materials and van der Waals heterostructures.

## Acknowledgments

ChatGPT (OpenAI; GPT-5.4 and GPT-5.5 Thinking) was used to assist with language editing, manuscript restructuring, preparation and checking of Python scripts for the additional numerical analyses, and refinement of schematic figures. All numerical calculations were executed and independently verified by the author. The author also verified the scientific content, interpretation, references, and conclusions and assumes full responsibility for the manuscript.

# Supplementary

## S.1 Geometric size-scaling of the NOT and AND gates

To complement the disorder and operating-window tests presented in the main text, we examined how the logical response depends on selected geometric dimensions of the NOT and AND nanostructures. Rather than attempting to define an atomically sharp universal minimum device size, we varied the characteristic control-patch lengths and branch widths around the reference geometries and recalculated the full truth table in each case. As in the main text, the performance metric was the minimum logical margin over all input states.

| | | | | |
|---|---|---|---|---|
| **NOT** | Total device length | $L$ | 60 | $0 \leq x \leq L$ |
| **NOT** | Input-region width | $W_{\mathrm{in}}$ | 22 | $-11 \leq y \leq 11$ |
| **NOT** | Splitter position | $x_{\mathrm{split}}$ | 20 | Input region ends and two output branches begin |
| **NOT** | Width of each output branch | $W_{\mathrm{branch}}$ | 10 | Upper: $2 \leq y \leq 12$; lower: $-12 \leq y \leq -2$ |
| **NOT** | Gap between output branches | $G_{\mathrm{branch}}$ | 4 | -2 < y < 2 |
| **NOT** | Common patch start | $x_{\mathrm{patch}}^{(0)}$ | 20 | Both complementary patches |
| **NOT** | Common patch end | $x_{\mathrm{patch}}^{(1)}$ | 34 | Both complementary patches |
| **NOT** | Common patch length | $L_{\mathrm{patch}}$ | 14 | $x_{\mathrm{patch}}^{(1)} - x_{\mathrm{patch}}^{(0)}$ |

| NOT | Upper patch region | $\Omega_{A,\mathrm{upper}}$ | — | $20 \le x \le 34, \quad 2 \le y \le 12$ |
|---|---|---|---|---|
| NOT | Lower patch region | $\Omega_{A,\mathrm{lower}}$ | — | $20 \le x \le 34, \quad -12 \le y \le -2$ |
| AND | Total device length | $L$ | 76 | $0 \le x \le L$ |
| AND | Stage-A splitter position | $x_A$ | 20 | Input region ends; upper and lower sectors begin |
| AND | Stage-B splitter position | $x_B$ | 46 | Upper sector splits into high and middle rails |
| AND | Total input width | $W_{\mathrm{tot}}$ | 32 | $-16 \le y \le 16$ |
| AND | Upper-sector width | $W_{\mathrm{upper}}$ | 20 | $-4 \le y \le 16$ |
| AND | Width of each final upper rail | $W_{\mathrm{sub}}$ | 8 | Middle: $-4 \le y \le 4$; high: $8 \le y \le 16$ |
| AND | Lower-branch width | $W_{\mathrm{lower}}$ | 8 | $-16 \le y \le -8$ |
| AND | Gap between high and middle rails | $G_{\mathrm{sub}}$ | 4 | $4 < y < 8$ |
| AND | Gap between upper and lower sectors | $G_{\mathrm{ul}}$ | 4 | $-8 < y < -4$ |
| AND | Stage-A patch interval | $[x_A^{(0)}, x_A^{(1)}]$ | [24,34] | $L_A = 10$ |
| AND | Stage-A upper patch | $\Omega_{A,\mathrm{upper}}$ | — | $24 \le x \le 34, \quad -4 \le y \le 16$ |
| AND | Stage-A lower patch | $\Omega_{A,\mathrm{lower}}$ | — | $24 \le x \le 34, \quad -16 \le y \le -8$ |
| AND | Stage-B patch interval | $[x_B^{(0)}, x_B^{(1)}]$ | [52,64] | $L_B = 12$ |
| AND | Stage-B high patch | $\Omega_{B,\mathrm{high}}$ | — | $52 \le x \le 64, \quad 8 \le y \le 16$ |
| AND | Stage-B middle patch | $\Omega_{B,\mathrm{mid}}$ | — | $52 \le x \le 64, \quad -4 \le y \le 4$ |

**Supplementary Table S1.** Reference geometric parameters of the NOT and AND nanostructures used in the truth-table, current-map, disorder, operating-window, and size-scaling calculations. All coordinates and lengths are expressed in units of the lattice constant a=1.

The quoted coordinate bounds define the continuous shape functions from which the discrete honeycomb devices are generated. Consequently, the precise atomic boundary follows the row-resolved lattice discretization. Dangling sites are removed before the semi-infinite leads are attached.
For the NOT gate, the common longitudinal length $L_{\mathrm{patch}}$ of the upper and lower complementary branch-selective patches was varied over the range $L_{\mathrm{patch}} = 1, \ldots, 14$, while the branch width was varied independently. For every input state, exactly one of the two patches was active. Over the tested patch-length range, the NOT gate remained correctly classified for all cases, confirming that the reference inverter is not tied to a single finely tuned longitudinal patch dimension. By contrast, the branch width proved more restrictive: in the adopted row-resolved honeycomb geometry, stable

NOT operation was recovered for several sufficiently wide and row-compatible branches, while narrower or geometrically unfavorable widths could suppress the output discrimination and lead to loss of correct classification. This indicates that the practical lower-size constraint for the NOT gate is governed more strongly by the branch width than by the common patch length.

For the AND gate, we varied the lengths of the stage-A and stage-B control patches separately, as well as the widths of the upper subbranches and the lower complementary branch. The correct AND truth table was retained throughout all tested values of the stage-A and stage-B patch lengths. However, the logical margin showed a stronger dependence on the stage-A length, with the reference value $L_A = 10$ lying near a high-margin operating point. In contrast, the stage-B length had only a weak influence on the margin over the tested range. A similar trend was observed for the branch widths: the width of the upper subbranches had only a minor effect, whereas the lower-branch width produced a more noticeable but still non-destructive variation of the margin. Thus, within the tested geometric window, the AND gate remains functionally robust, but its stage-A geometry is the most important design parameter.

The non-monotonic behavior visible in several scans is expected in a finite coherent nanostructure with a discrete honeycomb geometry, where row commensurability and mode matching can modify the transmission balance in an oscillatory way. The purpose of these scans is therefore not to extract a unique geometry-independent minimum size, but rather to demonstrate that the chosen reference devices are embedded in a finite and practically accessible geometric design window.

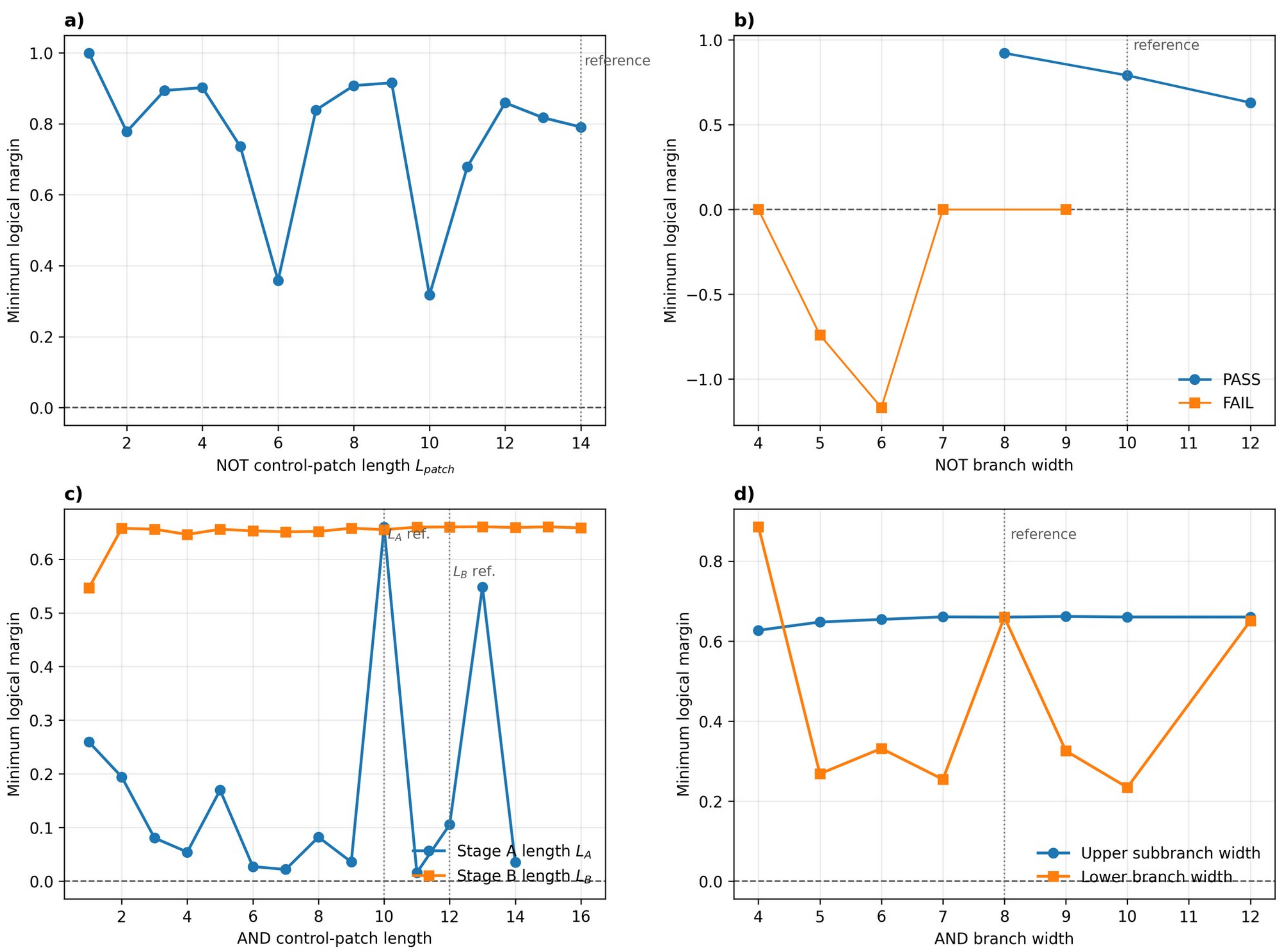

**Supplementary Fig. S1**. Geometric size-scaling of the NOT and AND gates. (a) Minimum logical margin of the NOT gate as a function of the common longitudinal length $L_{\text{patch}}$ of the two complementary upper and lower control patches. (b) Minimum logical margin of the NOT gate as a function of branch width; widths that reproduce the correct truth table are marked as PASS, while widths that fail are marked as FAIL. (c) Minimum logical margin of the AND gate as a function of the stage-A and stage-B control-patch lengths, $\mathrm{L}_A$ and $\mathrm{L}_B$. (d) Minimum logical margin of the AND gate as a function of the upper-subbranch width and the lower-branch width. Vertical dotted lines indicate the reference geometry used in the main text. The horizontal dashed line marks zero margin, i.e. the boundary between correct and incorrect logical classification.